\documentclass[12pt]{article}
\usepackage{amsmath}
\usepackage{amssymb}
\usepackage{bm}
\usepackage{graphicx}
\usepackage{epstopdf}
\usepackage{setspace}
\usepackage{color}
\usepackage{makecell}
\usepackage{multirow}
\usepackage{booktabs}
\usepackage{authblk}
\usepackage{textcomp}
\usepackage{hyperref}
\usepackage[title]{appendix}
\usepackage{longtable}
\usepackage{adjustbox}

\def\acknowledgments{\vskip12pt\noindent{\bf
Acknowledgments\vrule depth 6pt
width0pt\relax}\\*\noindent\ignorespaces}

\newcommand{\appropto}{\mathrel{\vcenter{
      \offinterlineskip\halign{\hfil$##$\cr
        \propto\cr\noalign{\kern2pt}\sim\cr\noalign{\kern-2pt}}}}}

\def\@cite#1#2{{#1\if@tempswa , #2\fi}}

\def\@citex[#1]#2{%
  \let\@citea\@empty
  \@cite{\@for\@citeb:=#2\do
    {\@citea\def\@citea{,\penalty\@m\ }%
     \edef\@citeb{\expandafter\@firstofone\@citeb}%
     \if@filesw\immediate\write\@auxout{\string\citation{\@citeb}}\fi
     \@ifundefined{b@\@citeb}{\mbox{\reset@font\bfseries ?}%
       \G@refundefinedtrue
       \@latex@warning
         {Citation `\@citeb' on page \thepage \space undefined}}%
       {
\csname b@\@citeb\endcsname
}}}{#1}}

\newdimen\bibindent
\def\newblock{\hskip .11em\@plus.33em\@minus.07em}

\RequirePackage{apacite}
\let\cite\shortcite 
\let\citeA\shortciteA 

\begin{document}

\title{When does vapor pressure deficit drive or reduce
  evapotranspiration?}

\author[1]{Adam Massmann\thanks{Corresponding author: akm2203@columbia.edu}}
\author[1]{Pierre Gentine}
\author[1,2]{Changjie Lin}

\affil[1]{Department of Earth and Environmental Engineering, Columbia University, New York, NY 10027}
\affil[2]{State Key Laboratory of Hydroscience and Engineering, Department of Hydraulic
  Engineering, Tsinghua University, Beijing, CN 100084}

\maketitle

\begin{abstract}
Increasing vapor pressure deficit (VPD) increases atmospheric demand
for water. While increased evapotranspiration (ET) in response to
increased atmospheric demand seems intuitive, plants are capable of
reducing ET in response to increased VPD by closing their stomata. We
examine which effect dominates the response to increasing VPD:
atmospheric demand and increases in ET, or plant response (stomata
closure) and decreases in ET. We use Penman-Monteith, combined with
semi-empirical optimal stomatal regulation theory and underlying water
use efficiency, to develop a theoretical framework for assessing ET
response to VPD. The theory suggests that depending on the environment
and plant characteristics, ET response to increasing VPD can vary from
strongly decreasing to increasing, highlighting the diversity of plant
water regulation strategies.

The ET response varies due to: 1) climate, with tropical and temperate
climates more likely to exhibit a positive ET response to increasing
VPD than boreal and arctic climates; 2) photosynthesis strategy, with
C3 plants more likely to exhibit a positive ET response than C4
plants; and 3) plant type, with crops more likely to exhibit a
positive ET response, and shrubs and gymniosperm trees more likely to
exhibit a negative ET response. These results, derived from previous
literature connecting plant parameters to plant and climate
characteristics, highlight the utility of our simplified framework for
understanding complex land atmosphere systems in terms of idealized
scenarios in which ET responds to VPD only. This response is otherwise
challenging to assess in an environment where many processes co-evolve
together.
\end{abstract}

\onehalfspacing

\section{Introduction}

Vapor pressure deficit (VPD) is expected to rise over continents in
the future due to the combination of increased temperature and,
depending on region, decreased relative humidity
\cite{Byrne_2013}. Increases in VPD increase the atmospheric demand
for evapotranspirated water \cite{Penman_1948, Monteith_1965}, but
also reduce stomatal conductance through stomatal closure
\cite{Rawson1977, Leuning_1990, Mott2007, Damour2010,
MEDLYN_2011}. Understanding the net evapotranspiration (ET) response
to these two opposing effects of changes in VPD is crucial for
assessing the impact of environmental VPD perturbations on the water
cycle.

The opposing effects of increased atmospheric demand and higher
stomatal closure lead to two possible perspectives for how ET responds
to shifts in VPD. The first, a hydrometeorological perspective, is
that higher VPD increases atmospheric demand for water from the land
surface, and this drives an increase in ET
\cite{Penman_1948}. However, plants' stomata have evolved to
optimally regulate the exchange of water and carbon, and tend to
partially close in response to increased atmospheric dryness
\cite{Farquhar_1978, Ball_1987, Leuning_1990, Katul_2009,
MEDLYN_2011}. This leads to a plant physiology perspective, in which
an increase in VPD may actually correspond to a decrease in ET because
of stomatal closure \cite<e.g.>[]{Rigden_2017}.  In other words, the
question ``When does VPD drive or reduce ET?'' can be related to
whether plant regulation or atmospheric demand dominates the ET
response.

The ET response to changes in VPD alters water partitioning between
the soil and atmosphere. If ecosystem plant response reduces ET with
atmospheric drying then soil moisture will be better conserved. This
represents a sensible evolutionary strategy to cope with aridity: save
water for periods when atmospheric demand for water is relatively low,
and atmospheric carbon can be accessed with a relatively smaller cost
in water loss. If instead stomata were fully passive \cite<similar to
soil pores, e.g. >[]{Or_2013}, increased atmospheric aridity would
strongly reduce soil moisture \cite{Berg_2017}. This could further
increase aridity as low soil moisture levels increases the Bowen
ratio, leading to increased temperature and atmospheric drying
\cite{Bouchet_1963, Morton_1965, Brutsaert_1999, Ozdogan_2006,
  Salvucci_2013, Gentine_2016, Berg_2016, Zhou_2019}. Therefore,
passive regulation and a lack of soil moisture conservation does not
seem to be a sensible strategy for plants from an evolutionary
standpoint. This simplified logic explains generally why plants
evolved to respond to VPD, but also excludes many details and special
cases \cite<e.g. plant to plant interaction, thermal regulation with
transpiration, and highly specialized photosynthesis strategies like
Crassulacean acid metabolism photosynthesis; see>[]{Brooker_2006, Lin_2017,
  Cushman_2001}.

We can use intuition about plant water conservation strategy to
hypothesize about ET response to changes in VPD. Plants and ecosystems
that evolved to conserve water, such as arid shrubs, should be more
likely to reduce ET with increasing VPD, and plants that have evolved
or have been engineered to prioritize carbon gain over water
conservation, such as crops, will be more likely to increase ET with
increasing VPD. Atmospheric conditions must matter as well. At the
ecosystem scale, there are limits to plant water conservation
strategies. As atmospheric demand for water (VPD) increases,
ecosystems may begin to reach their water conservation limits and
might not be able to entirely limit ET flux to the atmosphere. At this
stage any further increase in VPD will most likely drive a (limited)
increase in ET, because the increase in atmospheric demand for water
overwhelms the limited plant response to conserve water.

The objective of the present manuscript is to use reasonable
approximations established in prior research as a tool to develop a
framework for understanding plant responses to atmospheric drying and
the VPD dependence of ET while keeping other variables fixed. This
framework will aid interpretation of observations, full complexity
models, and facilitate the disentanglement of complex land-atmosphere
feedbacks. In particular, our approach has applications for
understanding climate change impacts, given expected increases in VPD
with rising temperature.  In the past, similar simplified approaches
were used to understand interactions between stomatal conductance,
evapotranspiration and the environment \cite<e.g.,>[]{Jarvis_1984,
Jarvis_1986, Mcnaughton_1991}. However these researchers did not
explore explicitly the sensitivity of ET to VPD, including VPD's
effect on stomatal conductance and plant function.

Approaching the problem of ET response to VPD is aided by recent
results drastically improving our understanding of VPD's impact on
physiology, especially at the leaf level. \citeA{MEDLYN_2011}
developed a model for leaf-scale stomatal conductance ($g_s$),
including VPD response, by combining an optimal photosynthesis theory
\cite{Cowan_1977, Katul_2009, Katul_2010} with an empirical approach,
and extended use of this model to the ecosystem scale in
\citeA{Medlyn_2017}. Additionally, \citeA{Zhou_2014} demonstrated that
a quantity underlying water use efficiency $\left(uWUE = \frac{GPP\;
\sqrt{VPD}}{ET}\right)$ properly captures a constant relationship
between GPP, ET, and VPD over a diurnal cycle at the ecosystem
scale. uWUE is also relatively well conserved in the growing season
across space and time, within a plant functional type (PFT)
\cite{Zhou_2015}.  While stomatal conductance parameterizations and
uWUE greatly simplify complex plant physiological processes, they
 still capture ecosystem behavior for vegetated surfaces
\cite{Medlyn_2017, Zhou_2014}, and are useful tools to transparently
develop intuition for the behavior of complex land-atmosphere systems.

In this manuscript, we leverage uWUE and recent developments in
stomatal conductance parameterizations \cite{MEDLYN_2011} with a
Penman-Monteith framework \cite<hereafter PM,>[]{Penman_1948,
Monteith_1965} to derive the theoretical one-way response of ET to VPD
with other environmental variables properly controlled for, i.e. we
develop a framework for evaluating the partial derivative of ET with
respect to VPD. It is useful to disconnect the impact of VPD from
other variables as VPD is known to increase dramatically with future
climate and limit ET more than soil moisture \cite{Novick_2016}, but
disentangling VPD effects from soil moisture effects has been
difficult in previous research, given their co-variability
\cite{Lin_2018, Zhou_2019}.  We are
able to disentangle the effect of VPD on ET because we explicitly
include VPD's full effect on stomatal conductance, including its
impact on photosynthesis. The use of an energy balance framework (PM)
allows us to include in our analysis the effects of the energy cost of
evaporating water from a surface, which is an important factor in the
natural environment, compared to prescribed in situ environmental
conditions. This is relevant because previous research focusing on the
leaf scale \cite{Rawson1977, Turner1984, Oren1999, Damour2010,
Mott2013} does not consistently or analogously include the energetic
constraints on evapotranspiration under which ecosystems operate. The
leaf-scale results agree that stomatal conductance decreases in
response to VPD \cite{Oren1999, Damour2010}, which we expect to be
true for an ecosystem as well. However leaf scale results also
indicate that transpiration usually increases with increasing VPD in a
concave downward shape \cite<e.g.,>[]{Rawson1977, Turner1984,
Mott2013}, which may not be true for ET at the ecosystem scale once
energetic constraints on ET are included in the analysis.  Our
approach allows us to estimate the expected ET response to VPD at the
ecosystem scale, including the effects of surface energy balance
constraints, and assess how ecosystem ET response to VPD deviates from
previous leaf scale analyses.

This manuscript presents the range of possible ecosystem-scale ET
responses to VPD, given parameters previously established in peer
reviewed literature. Additionally, we explore the sensitivity of the
ET-VPD relationship to stomatal model and framework choice,
highlighting the importance of: 1) future research on stomatal
conductance and ecosystem scale modeling, and 2) thoughtful selection
of photosynthesis and stomatal conductance models in more
sophisticated land surface and earth system models.

\section{Methods}
\label{methods}
The Penman-Monteith equation \cite<hereafter PM,>[]{Penman_1948,
  Monteith_1965} estimates ET as a function of observable atmospheric
variables and surface conductances:
  \begin{equation}
    \label{orig_pen}
    ET = \frac{\Delta R_{net} + g_a \rho_a c_p VPD}{\Delta + \gamma(1 + \frac{g_a}{g_s})},
  \end{equation}
  where $\Delta$ is the change in saturation vapor pressure with
  temperature, given by Clausius-Clapeyron
  ($\frac{d \; e_s}{d \; T}$), $R_{net}$ is the net radiation minus
  ground heat flux, $g_a$ is aerodynamic conductance, $\rho_a$ is air
  density, $c_p$ is specific heat of air at constant pressure,
  $\gamma$ is the psychometric constant, and $g_s$ is the stomatal
  conductance (Table \ref{definitions}). The issue with PM is that
  $g_s$ is a function of carbon uptake, which is has a strong
  functionally relation to ET through stomatal function. Therefore,
  when PM is formulated in terms of $g_s$, it is an implicit function
  of ET itself rather than an explicit function of ET. Here we will
  derive a new form of PM in which ET is an explicit function of plant
  parameters and environmental conditions, and use it to assess the
  ecosystem scale response to VPD (by taking a partial derivative).

\begin{table}
  \caption{Definition of symbols and variables, with citation for how
    values are calculated, if applicable.}
  \label{definitions}
  \centering \footnotesize
  \begin{adjustbox}{center=15cm}
    \begin{tabular}{l c c c}
    \hline
    Variable & Description & Units & Citation \\
    \hline
    $e_s$  & saturation vapor pressure & Pa  & - \\
    $T$  & temperature  & K & - \\
    $P$  & pressure & Pa  & - \\
    $\Delta$  & $\frac{\partial e_s}{\partial T}$ & Pa K$^{-1}$ & - \\
    $R_{net}$  & net radiation at land surface minus ground heat flux & W m$^{-2}$   & - \\
    $g_a$  & aerodynamic conductance & m s$^{-1}$  & \citeA{Shuttleworth_2012} \\
    $\rho_a$  & air density & kg m$^{-3}$  & - \\
    $c_p$  & specific heat capacity of air at constant pressure & J K$^{-1}$ kg$^{-1}$ & - \\
    $VPD$  & vapor pressure deficit & Pa  & - \\
    $\gamma$  & psychometric constant & Pa K$^{-1}$   & - \\
    $g_{s}$  &  stomatal conductance & m s$^{-1}$
                                   & \citeA{Medlyn_2017} \\
    $g_{1}$  & ecosystem-scale slope parameter & Pa$^{0.5}$ & \citeA{Medlyn_2017} \\
    $R$ & universal gas constant & J mol$^{-1}$ K$^{-1}$ & - \\
    $R_{air}$ & gas constant of air & J  K$^{-1}$ kg$^{-1}$ & - \\
    $\sigma$ & uncertainty parameter & -& - \\
    $c_a$ & CO$_2$ concentration & $\mu$ mol CO$_2$ mol$^{-1}$ air& - \\
    $\lambda = \frac{\partial \; transpiration}{\partial\; CO_2\; assimilation}$ & marginal water cost of leaf carbon & mol H$_2$O mol$^{-1}$ CO$_2$ & - \\
    $\Gamma$ & CO$_2$ compensation point & - & - \\
    $\Gamma^*$ & CO$_2$ compensation point without dark respiration &
                                                                      - & - \\
    GPP & gross primary production & $\mu$-mol C s$^{-1}$ m$^{-2}$ & -
    \\
    ET & evapotranspiration & W m$^{-2}$ & - \\
   uWUE & underlying water use efficiency & $\mu$-mol C Pa$^{0.5}$
                                            J$^{-1}$ ET  &
                                                           \citeA{Zhou_2015} \\
    \hline
  \end{tabular}
\end{adjustbox}
\end{table}

\citeA{MEDLYN_2011} developed a model for stomatal conductance ($g_s$)
by combining an optimal photosynthesis theory \cite{Cowan_1977} with
an empirical approach, which describes the dependence of $g_s$ to
VPD. They also extended this model to the ecosystem scale
\cite{Medlyn_2017}:

  \begin{equation}
    g_s = \frac{R \,T}{P} \; 1.6 \left(1 + \frac{g_1}{\sqrt{VPD}}\right) \frac{GPP}{c_a},
    \label{medlyn}
  \end{equation}

  where $g_{1}$ is a VPD ``slope'' parameter, GPP is the ecosystem
  scale gross primary production, and $c_a$ is the atmospheric CO$_2$
  concentration at the canopy. \cite{MEDLYN_2011} relate the slope
  parameter ($g_{1}$) to physical parameters as:

  \begin{equation}
    g_{1}  \propto \sqrt{\frac{3 \, \Gamma^* \,
        \lambda}{1.6}},
    \label{slope}
  \end{equation}

  where g$_1$ is presented in units of $\sqrt{\text{mol fraction}}$ for
  convenience (which can be converted to $\sqrt{Pa}$ using the ideal
  gas law and ambient pressure), $\Gamma^*$ is the CO$_2$ compensation
  point for photosynthesis (without dark respiration), and $\lambda$
  is the marginal water cost of leaf carbon
  ($\frac{\partial \; \text{transpiration}}{\partial \; CO_2
    assimilation}$) \cite{Farquhar_1980, Katul_2009}. So, $g_{1}$ is a
  leaf-scale term reflecting the trade-off of water for carbon
  uptake. The higher $g_{1}$, the more open the stomata and the more
  they release water in exchange for carbon.

While Eq. (\ref{medlyn}) can be used in PM (Eq. (1)), it will make
analytical work with the function intractable because GPP is
functionally related to ET itself. Additionally, a perturbation to VPD
should induce a physiological plant response that will alter GPP and
cause an indirect change in stomatal conductance, in addition to the
direct effect of VPD in Eq. (\ref{medlyn}). Therefore, in order to
derive the full plant response of ET to VPD, we must account for the
functional relationship between GPP, ET, and VPD, and its effect on
stomatal conductance. We can use aforementioned semi-empirical results
of \citeA{Zhou_2015}, which were inspired by optimal photosynthesis
theory, as a tool to approach this problem. \citeA{Zhou_2015}, showed
that underlying Water Use Efficiency (uWUE):

  \begin{equation}
    uWUE = \frac{GPP \cdot \sqrt{VPD}}{ET}
    \label{uwue}
  \end{equation}

  is relatively constant across time within a plant functional type,
  and correctly captures a constant relationship between GPP, ET and
  VPD over a diurnal cycle during the growing season
  \cite{Zhou_2014}. The theoretical derivation of the square root VPD
  dependence in uWUE leverages the same assumptions used in
  \citeA{MEDLYN_2011} to derive the square-root VPD dependence of the
  stomatal conductance model (Eq.  (\ref{medlyn})).  We can use uWUE
  to remove the $GPP$ dependence in $g_s$ in a way that makes PM
  analytically tractable:

  \begin{equation}
    g_s = \frac{R \, T}{P} 1.6 \left(1 + \frac{g_1}{\sqrt{VPD}}\right) \frac{uWUE \; ET}{c_a \; \sqrt{VPD}}.
    \label{new_g_s}
  \end{equation}

Plugging Eq. (\ref{new_g_s}) into Eq. (\ref{orig_pen}) and
rearranging gives a new explicit expression for PM, in which
dependence on $GPP$ is removed:

  \begin{equation}
    ET = \frac{\Delta R_{net} + \frac{g_a\; P}{T} \left( \frac{ c_p VPD}{R_{air}} -  \frac{\gamma c_a \sqrt{VPD} }{ R \; 1.6\; \text{ uWUE } (1 + \frac{g_1}{\sqrt{VPD}})} \right) }{ \Delta + \gamma}
    \label{et}
  \end{equation}

  By accounting for photosynthesis changes in ecosystem conductance,
  with Eq. (\ref{et}) we have used recent results
  \cite{MEDLYN_2011, Zhou_2014, Zhou_2015, Medlyn_2017} to derive
  ET explicitly as function of environmental variables and two
  plant-specific constants, the slope parameter ($g_1$), and uWUE. For
  the first time we have removed the implicit dependence of ET on
  itself through the stomatal conductance term, and we have also
  replaced the added complexity of a stomatal conductance reduction
  factor and a photosynthesis model with a single parameter (uWUE). Both
  the plant parameters reflect water conservation strategy. The
  slope parameter is related to the willingness of stomata to trade
  water for CO$_2$ and to keep stomata open (carbon cost in terms of
  water)). uWUE is a semi-empirical ecosystem-scale constant related
  to how WUE changes with VPD (specifically $VPD^{-1/2}$). It is also
  roughly proportional to physical constants:

\[uWUE \appropto \sqrt{\frac{c_a - \Gamma}{1.6 \lambda}},\]

where $\Gamma$ is the CO$_2$ compensation point \cite<Eq. (5)
in>[]{Zhou_2014}. So uWUE is related to atmospheric CO$_2$
concentration and compensation point, and is inversely proportional to
the marginal water cost of leaf carbon ($\lambda$). This relationship
with $\lambda$ ($\propto \lambda^{-1/2}$) is important as it is the inverse of g$_1$'s
relationship with $\lambda$ ($\propto \sqrt{\lambda}$). So, uWUE
should increase as $\lambda$ decreases, and g$_1$ should decrease as
$\lambda$ decreases.

With Eq. (\ref{et}) we can take the partial derivative of ET with respect
to VPD to understand whether VPD drives or reduces ecosystem-scale ET:

  \begin{equation}
    \frac{\partial \;  ET}{\partial \; VPD} = \frac{2\; g_a \;
      P}{T(\Delta + \gamma)}   \left(\frac{ c_p}{R_{air}} -
      \frac{\gamma c_a }{1.6 \; R\; \; \text{ uWUE }} \left(
        \frac{2 g_1 + \sqrt{VPD}}{2 (g_1 + \sqrt{VPD})^2}\right)
    \right),
    \label{d_et}
  \end{equation}

  providing analytical framework for ecosystem response to atmospheric
  demand with environmental conditions held fixed. There are a few
  subtleties to taking the derivative in Eq. (\ref{d_et}): $\Delta$
  ($\frac{d e_{s}}{d T}$) and $VPD$ are functionally related, so while
  taking the derivative we evaluate
  $\frac{\partial \; ET}{\partial \; VPD} = \frac{\partial \; ET}
  {\partial \; e_s} \frac{\partial \; e_s}{\partial \; VPD}
  \Big|_{\text{RH fixed}} + \frac{\partial \; ET}{\partial \; RH}
  \frac{\partial \; RH}{\partial \; VPD} \Big|_{\text{$e_s$
      fixed}}$. $RH$ and $e_s$ are assumed to be approximately
  independent, which is supported by data (see supplementary material).

  This derivation relied either implicitly or explicitly on several
  assumptions. First, we assume that VPD at the leaf surface is the
  same as VPD at measurement height; physically this implies that
  leaves are perfectly coupled to the atmosphere. In reality, for some
  conditions and plant types the leaves can become decoupled from the
  boundary layer \cite{De_2017, Medlyn_2017}. Therefore, our
  derivation will be most applicable in times like the growing season
  (when we also expect uWUE to be most valid), when relatively high
  insolation induces instability and convective boundary layers, and
  we would expect the surface to be generally well coupled. An
  additional assumption in the formulation of uWUE \cite{Zhou_2014,
    Zhou_2015} and the stomatal conductance model of
  \citeA{Medlyn_2017} is that direct soil evaporation contributions to
  ET remain small relative to transpiration. Again, this should be
  more true during the growing season. The ratio of evaporation to
  transpiration may increase immediately after rainfall events due to
  high soil moisture, ponding, and interception, but VPD is generally
  low anyways during these times. However, some plant types allow for
  systematically larger contributions of evaporation in ET,
  particularly those with sparse canopies and smaller relative amounts
  of transpiration. We therefore might expect that the theory will be
  most applicable to forest PFTs, which will be most strongly coupled
  to the boundary layer due to larger surface roughness, and will also
  generally have the highest ratios of transpiration to
  evaporation. Finally, we assume that $g_1$ and uWUE are constant
  with respect to the conceptual VPD perturbation. Both quantities
  have been shown to be relatively constant with respect to changes in
  VPD \cite{Franks_2017, Zhou_2014}. These parameters will however
  vary with plant species and characteristics \cite<e.g. wood density,
  >[]{Lin_2015}, as well as environmental conditions including soil
  water content and temperature \cite{Lin_2015,
    Manzoni2013}. Exploring possible soil moisture (in)dependence of
  the plant parameters (g$_1$ and uWUE) is particularly interesting
  because soil moisture only enters the partial derivative directly
  through these plant parameters. If the plant parameters are weak
  functions of soil moisture then the theory can be directly applied
  to a broader range of conceptual VPD scenarios, including observed
  compound events between high VPD and low soil moisture
  \cite{Zhou_2019}. Supplementary material for this manuscript
  explores the soil moisture dependence of the plant parameters, but
  excessive noise and inconsistencies preclude conclusions about the
  nature of any dependence. We provide it in case it is of interest to
  the reader, and to motivate future research.

  Because Eq. (\ref{d_et}) is a partial derivative, its utility is as
  a conceptual model for the change in ET in response to VPD with all
  other variables held fixed. This provides a useful tool for
  identifying the effect of VPD on ET through atmospheric demand and
  plant response, and allows a practitioner to disentangle complicated
  feedbacks when many quantities co-vary. However users of
  Eq. (\ref{d_et}) should take care that their interpretation matches
  the assumptions inherent in a partial derivative. For example,
  results will only be as valid as the assumption that g$_1$ and uWUE
  (and by extension $\lambda$) are fixed with respect to the user's
  conceptual change in VPD. Care must also be taken with possible
  indirect effects associated with a change in VPD: for example, a
  change in ET induced by a change in VPD can cause a change in
  surface temperature, which would drive a change in net
  radiation. These types of indirect effects and feedbacks are not
  considered in Eq. (\ref{d_et}): temperature (a variable) is
  mathematically fixed.

\subsection{Framing Eq. \ref{d_et} with previous research }

  To account for both the variability in plant parameters and the
  environment, we will systematically analyze how the ET response to
  VPD (Eq. (\ref{d_et})) varies. Eq. (\ref{d_et}) includes a
  ``sign'' term that determines the sign of the response in addition
  to magnitude:

\begin{equation}
  \label{sign}
  \frac{c_p}{R_{air}} - \frac{\gamma c_a }{1.6 \; R\; \text{ uWUE }} \left( \frac{2 g_1 + \sqrt{VPD}}{2 (g_1 + \sqrt{VPD})^2}\right),
\end{equation}

and a ``scaling'' term multiplying the ``sign'' term:

\begin{equation}
  \frac{g_a \; P}{T(\Delta + \gamma)}.
\end{equation}

In the ``sign term'' most of the quantities are relatively constant,
except for VPD and the plant parameters g$_1$ and uWUE. In the scaling
term, most of the terms are relatively constant with the exception of
aerodynamic conductance ($g_a$) and temperature (especially its effect
through $\Delta$). To determine the range of probable ET responses to
VPD we will systematically vary these parameters according to Table
\ref{param_varying}, while all other parameters are held fixed (Table
\ref{param_fixed}). Physical variables ($g_a$, $T$) and plant
physiological parameters (g$_1$, uWUE) are varied according to
literature-based expectations for a range of growing season conditions
and plant types \cite{Zhou_2015, Medlyn_2017}. Using this previous
literature we can connect the effect of varying plant parameters to
specific plant types and characteristics.

All code and data used in this analysis, including those used to
generate the figures and tables, are publicly available at
\url{https://github.com/massma/climate_et}.

\begin{table}
  \caption{Variable quantities in the ET response to VPD. Each value
    is varied to determine the effect of a range of expected plant and
    environmental conditions on ET response to VPD. A citation is
    provided for values of g$_1$ and uWUE, which are directly derived
    from previous literature. Conceptually, min. values are extracted
    from literature to correspond to approximately the 15th percentile
    of observed conditions during the growing season, med. values
    correspond to approximately the 50th percentile, and max. values
    correspond to approximately the 85th percentile. Values for $T$
    and g$_a$ are calculated from FLUXNET-2015 data (see supplemental
    material for description), rounding the 15th, 50th, and 85th
    percentile to the nearest 5$^o$C and 0.005 m/s, respectively.}
  \label{param_varying}
  \centering
  \footnotesize
  \begin{adjustbox}{center=15cm}
  \begin{tabular}{l c c c c c}
    \hline
    Symbol & units [units in citation] & min  & med & max
    & citation  \\
    \hline
g$_1$ & Pa$^{1/2}$ [kPa$^{1/2}$] & 63.25 [2.00] & 126.49 [4.00] & 189.74 [6.00] & Fig. 2, 7; \citeA{Medlyn_2017} \\
uWUE & $\mu$-mol C Pa$^{0.5}$ J$^{-1}$ ET [g C hPa$^{1/2}$ kg$^{-1}$ H$_2$O] & 2.33 [6.99] & 3.17 [9.52] & 4.01 [12.05] & Table 4; \citeA{Zhou_2015} \\
T & $^o$C & 10.00 & 20.00 & 30.00 & - \\
g$_a$ & m/s & 0.015 & 0.035 & 0.055 & - \\
    \hline
  \end{tabular}
  \end{adjustbox}
\end{table}

\begin{table}
  \caption{Quantities that are fixed in the ET response to VPD
    (relative to those in Table \ref{param_varying}).}
  \label{param_fixed}
  \centering
  \begin{tabular}{l c c}
    \hline
    Symbol & units & value \\
    \hline
P & Pa & 100000.00 \\
$\gamma$ & Pa K$^{-1}$ & 64.50 \\
R$_{air}$ & J  K$^{-1}$ kg$^{-1}$ & 288.00 \\
c$_a$ & $\mu$ mol CO$_2$ mol$^{-1}$ air & 400.00 \\
    \hline
  \end{tabular}
\end{table}

\section{Results}
\label{results}

\begin{figure}
  \centering \includegraphics{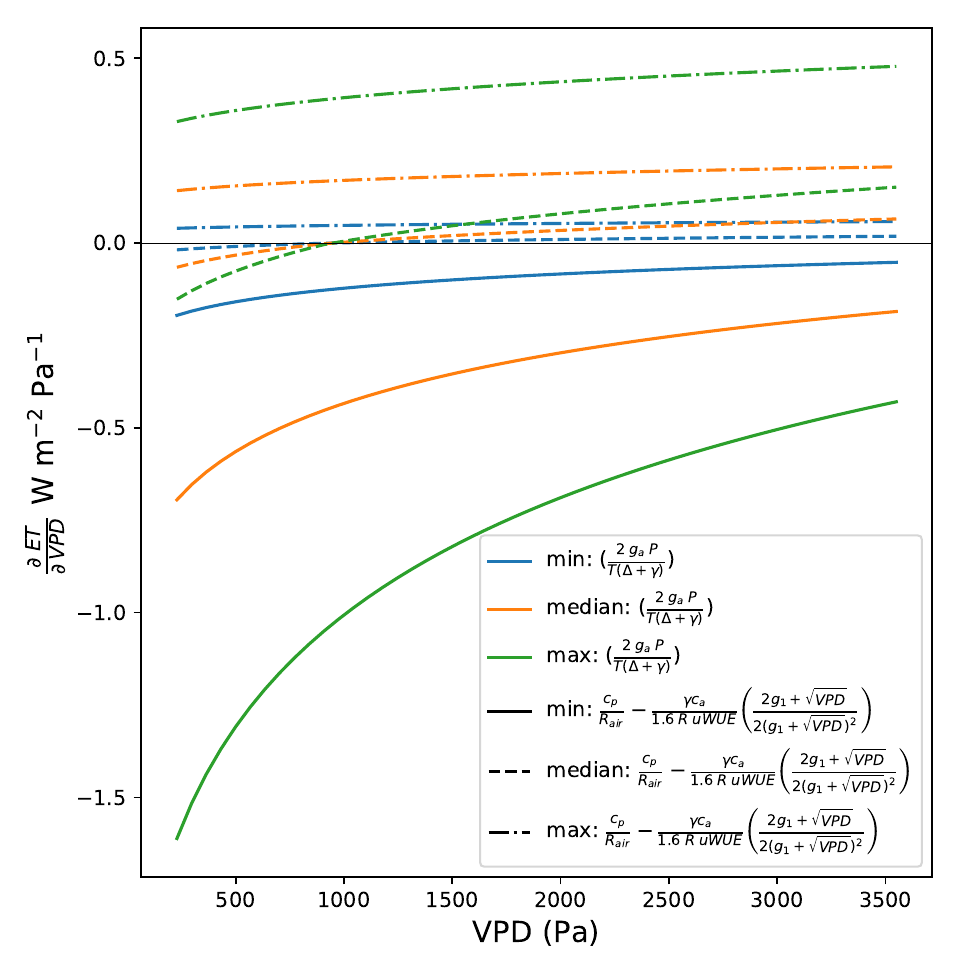}
  \caption{The functional form of $\frac{\partial \; ET}{\partial
      \; VPD}$ for minimum, median and maximum values of both the sign
    term and the scaling term.}
  \label{full}
\end{figure}

By varying four parameters (g$_a$, T, uWUE, g$_1$) at three different
values (Table \ref{param_varying}) we generate nine different values
for the scaling term, nine different curves (as a function of VPD) for
the sign term, and 81 different curves for the ET response to VPD
($\frac{\partial \; ET}{\partial \; VPD}$). To aid visualization we
can examine a subset of nine of these 81 curves, defined by the
minimum, median and maximum values for both the scaling and sign term
(Figure \ref{full}). The range of ET responses to VPD vary from those
where ET almost always decreases with increasing VPD (water
conservative), to those where ET almost always increases with
increasing VPD (water intensive). Additionally, for some parameters
whether ET will increase or decrease with increasing VPD depends on
atmospheric demand (Figure \ref{full}).

\begin{figure}
  \centering \includegraphics{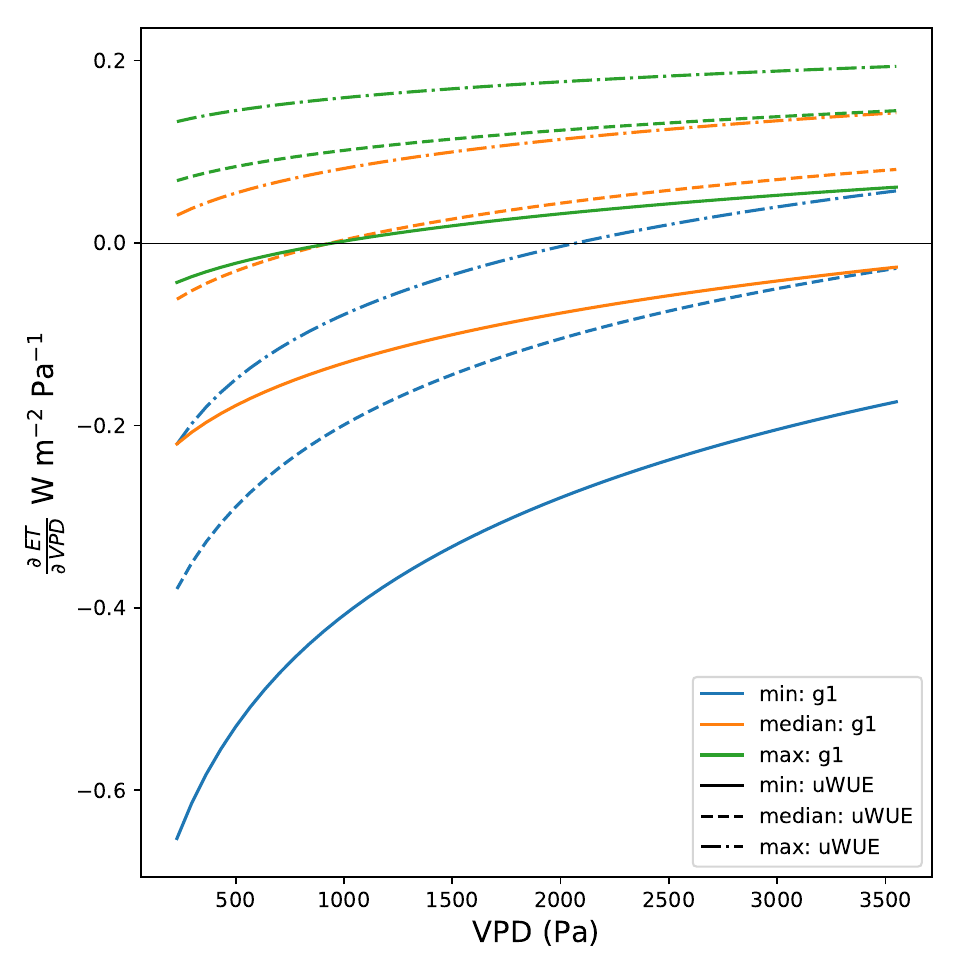}
  \caption{The functional form of $\frac{\partial \; ET}{\partial
      \; VPD}$ evaluated at the median value of the scaling term, for
    varying values of g$_1$ and uWUE as given in Table \ref{param_varying}.}
  \label{sign}
\end{figure}

Examining the sign term independent of the scaling term illuminates
the role of the two plant parameters, g$_1$ and uWUE, in determining
the degree to which the response is water conservative
($\frac{\partial \; ET}{\partial \; VPD} < 0$) or water intensive
($\frac{\partial \; ET}{\partial \; VPD} > 0$) (Figure \ref{sign})
. Higher g$_1$ and uWUE shift the curve towards increasing ET
responses with VPD (water intensive), and smaller g$_1$ values lead to
a larger VPD dependence of the response (i.e. ET response is a
stronger function of VPD). However care should be exercised when
interpreting the range of ET responses, because g$_1$ and uWUE should
generally be anti-correlated due to their dependencies on
$\lambda$. As $\lambda$ increases, g$_1$ should increase and uWUE
should decrease. Because high and low values of uWUE and g$_1$ were
determined independently from each other in previous literature, plant
types with anomalously high or low values for \textit{both} g$_1$ and
uWUE should be relatively rare. However, some plant types do exhibit
co-occurring high values for g$_1$ and uWUE. For example, C3 crops
have both the highest g$_1$ value in \citeA{Franks_2017}, and the
highest uWUE value in \citeA{Zhou_2015}.

\begin{figure}
  \centering \includegraphics{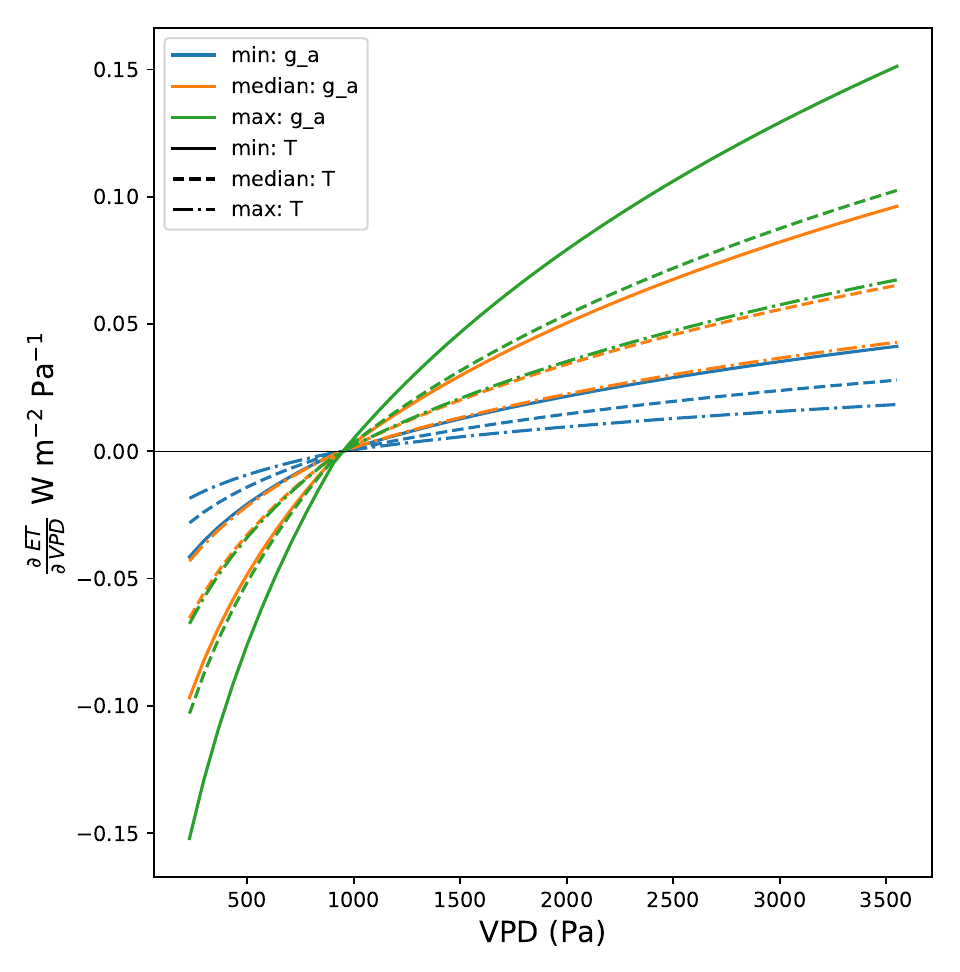}
  \caption{The functional form of $\frac{\partial \; ET}{\partial
      \; VPD}$ evaluated at the median value of the sign term, for
    varying values of g$_a$ and T as given in Table \ref{param_varying}.}
  \label{scaling}
\end{figure}

Examining the scaling term independent of the sign term shows how
aerodynamic conductance (communication between the atmosphere and the
surface) and temperature (controls the efficiency of energy conversion
to latent heat through $\Delta$), amplify or suppress the plant
response represented in the sign term (Figure \ref{scaling}). Both
lower temperatures and higher aerodynamic conductance lead to
amplified ET response to VPD, with the variability of aerodynamic
conductance resulting in a slightly higher variability of ET response
relative to temperature.

\section{Discussion}
\label{discussion}

Interpreting both the derivation and results of this manuscript hinges
on an understanding and appreciation for the usefulness of partial
derivatives for understanding the behavior of complex systems in a
simplified framework. Our results are a conceptual tool for answering
the question, ``What is the ET response to VPD with all other
quantities held fixed?'' This is useful, and we would argue is
critical, for understanding complex responses when many quantities are
varying simultaneously \cite{Zhou_2019}. If we cannot understand the
contribution of one term to observed variability, independent of other
confounding factors, there is little hope of disentangling and
understanding the relative role of many co-varying quantities in
determining observed variability.  Eq. (\ref{d_et}) explicitly
provides an estimate of the ET response to VPD, given assumptions
about ecosystem plant characteristics through the parameters g$_1$ and
uWUE. Eq. (\ref{d_et}) is generic to external environmental factors
and the timescale over which a VPD response acts, and has a
straightforward conceptual interpretation. It provides the ET
response, given other quantities and parameters held fixed, subject to
the assumptions outlined in Sect. \ref{methods}. So, while the plant
parameters g$_1$ and uWUE may vary with plant species and
environmental conditions like soil moisture, we can still assess the
ET response to VPD for a given ecosystem state, or for a given soil
moisture condition.

However, given that g$_1$ and uWUE can vary, and they both have a
large impact on the ET response (Figure \ref{full}, \ref{sign}), it is
useful to examine how these quantities vary in previous literature,
framed by our results on $\frac{\partial \; ET}{\partial \; VPD}$. In
\citeA{Zhou_2015} most plant types' uWUE is similar to our median
value, under which the sign of the ET response depends on
VPD. However, crops, shrubs, and grass are exceptions; they exhibit an
uWUE different than the median value in Table
\ref{param_varying}. Crops, which we expect to prioritize carbon
uptake over water conservation, have a higher uWUE value closer to our
maximum value. This higher uWUE results in a higher likelihood for an
increasing ET response to VPD, which matches intuition given that we
expect crops to keep stomata open for access to carbon, at the cost of
increased water loss during high VPD. Shrubs, and to a lesser extent
grass, have an uWUE closer to the minimum uWUE. For these plant types,
we would then expect a decrease of ET in response to VPD. It is
important to note that, while \citeA{Zhou_2015} did not examine the
role of soil moisture for within plant-type variability of uWUE, we
might expect some variation, especially in extreme cases when soil
water becomes a limiting factor (see supplementary material).

The ecosystem-scale results of \citeA{Medlyn_2017} did not display the
robust relationship between plant type and and plant parameter (g$_1$)
that was demonstrated by \citeA{Zhou_2015} for uWUE. Given that there
is a robust relationship between g$_1$ and plant type in the
leaf-scale results of \citeA{Lin_2015}, and the ecosystem-scale
results of \citeA{Zhou_2015} rely on the same underlying theory as in
\citeA{Medlyn_2017}, the ambiguous results in \citeA{Medlyn_2017}
could be due to noise in ecosystem scale observations, the
consequences of imposing a model structure in some of their g$_1$
estimations, or real ecosystem-scale within-plant type variability
(e.g. multiple species impacting the relationship within the
eddy-covariance footprint). These differences between leaf-scale
behavior and ecosystem scale behavior highlight the importance of
understanding how and why slope parameters show increased variability
at the ecosystem scale, as argued by \citeA{Medlyn_2017}. However, if
we assume that there is some real analogy between leaf scale results
in \citeA{Lin_2015} and ecosystem scale behavior, and that some of the
ambiguity in \citeA{Medlyn_2017} is due to model and/or observational
error, then we can use the relationships between g$_1$ and plant
characteristics defined by \citeA{Lin_2015} to frame the ET response
to VPD. This is subject to the strong caveat that we still do not
fully understand g$_1$ behavior at the ecosystem scale. Extrapolating
the results of \citeA{Lin_2015} (e.g. Fig. 2) to the ecosystem scale
would reveal the following relationships between g$_1$ and plant
types:

\begin{itemize}
  \item C3 plants would have a generally higher g$_1$ than C4 plants,
  \item Crops would have a larger g$_1$ than shrubs, grass, and
    angiosperm trees, which have a higher g$_1$ than gymnosperm
    trees.
  \item Tropical and temperate climates would generally be
    characterized by plants with a higher g$_1$ than arctic and boreal
    climates.
\end{itemize}
For all of the above relationships, a higher g$_1$ means a higher
likelihood of a positive ET response to increasing VPD (water
intensive), and a smaller VPD dependence of the ET response.

Our theoretical results highlight the variability of ET
response to VPD as a function of both climate  and plant
characteristics, especially water usage strategy. Whether an
ecosystem increases or decreases ET in response to VPD will vary
depending on vegetation and climate. Generalizations about ET response
to VPD therefore require thoughtful consideration of ecosystem
characteristics; statements such as ``ET increases with warming due to
increases in VPD'' may be false depending on the water conservation
strategy of a given ecosystem \cite{Lemordant_2018}.

Additionally, the sensitivity of ET response to plant parameters
highlights the importance of understanding and developing stomatal
conductance models and parameters for the ecosystem scale. In
particular, why and how g$_1$ behavior at the leaf scale is not
analogous to the ecosystem scale must be understood. While the
derivation here was explicit about the assumption of a constant g$_1$
term with respect to the VPD perturbation, many sophisticated land
surface models and earth system models employing stomatal conductance
models make similar assumptions about the stationarity of VPD slope
terms \cite{Niu_2011, Franks_2017, Rogers_2017, Lawrence_2019}. Many
models will assume a constant VPD slope term (e.g. g$_1$) within
a given plant functional type, and research derived from these models
often does not acknowledge limitations of this assumption, or the
difficulty in the literature with quantifying an ecosystem scale slope
parameter \cite{Medlyn_2017}. Given that something so fundamental as
the ET response to VPD varies strongly with g$_1$, we must invest more
in understanding g$_1$'s behavior at the ecosystem scale, and the
efficacy of models using constant slope terms for ecosystem-scale
fluxes. Fortunately, our framework, and the new approach of using
robust ecosystem-scale ratios (uWUE) to remove the implicit dependence
of stomatal conductance on ET, is flexible enough to be applied to any
stomatal conductance model that contains a dependence on GPP.

So far we have discussed the role of different parameters in
modulating the ET response. However, the form of the ET response is
imposed by our choice of stomatal conductance model. We now examine
how stomatal model choice, and specifically the exponent of the VPD
dependence in the stomatal conductance model, can impact the general
form of the ET-VPD relationship. There is a theoretical basis for the
square root VPD dependence in both the stomatal conductance model and
uWUE based on the assumption that stomata behave to maximize carbon
gain while minimizing water loss, which observations also generally
support \cite{Lloyd_1991, MEDLYN_2011, Lin_2015, Zhou_2014, Zhou_2015,
  Medlyn_2017}. However, some purely empirical results that fit the
exponent of the VPD dependence to data have shown that it may vary
slightly from 1/2, suggesting that stomata, as well as ecosystem-scale
quantities based on stomata theory, may not always function optimally
\cite{Zhou_2015, Lin_2018}. Specifically with regards to uWUE, one
would not expect that this ecosystem scale WUE quantity will respond
to VPD exactly analogously to stomata. Direct soil evaporation's
contributions to ET should shift the exponent of the VPD
dependence. The results from \citeA{Zhou_2015} corroborate this: they
found a mean empirically fit exponential VPD dependence of 0.55,
varying slightly from the theoretically optimal value of of 0.5 for
AmeriFlux sites. Results in \citeA{Lin_2018} also show variance in the
empirical exponent of the VPD dependence in the stomatal conductance
model, but interpretation of this variance is more difficult as
\citeA{Lin_2018} do not handle the GPP dependence of stomatal
conductance in a directly analogous manner to the optimal theory in
\citeA{MEDLYN_2011} and \citeA{Medlyn_2017}. Regardless, given that
these recent results on the relationship between VPD, GPP, and ET
\cite{MEDLYN_2011, Zhou_2014, Zhou_2015, Medlyn_2017} form the
backbone of our analysis and are what allowed us to derive an explicit
ET expression for the first time (Eq. (\ref{et})), we will analyze if
and how assumptions about the exponent of the VPD dependence impact
the shape of the ET-VPD dependence. This analysis is also important to
understand whether the choice of stomatal conductance model alters the
fundamental behavior of the ET-VPD relationships, as many commonly
used models utilize a VPD exponent other than the 1/2 suggested by
optimal theory \cite<e.g. >[ which uses an exponent of
1]{Leuning_1990}. It is useful to understand how the choice of model
imposes the functional form of the ET-VPD relationship, given its
fundamental importance for land-atmosphere interactions.

\subsection{Functional form of ET dependence on VPD and its relation
  to the VPD exponent}
\label{functional_form}

The results presented in Sect. \ref{results} indicates that for a
given uWUE and $g_1$, the ET dependence on VPD should be concave
upward. In other words, there should be some local minimum in ET at a
critical VPD$_{crit}$, assuming the scaling and plant terms
(e.g. aerodynamic conductance, $\Delta$, $g_1$ and uWUE) are held
fixed. This result warrants further investigation, because to our
knowledge no earlier work has derived the theoretical ecosystem-scale
relationship between ET and VPD in an energy balance framework, while
controlling for other environmental conditions. In particular there is
an apparent lack of consensus over whether the shape of the ET-VPD
curve should be concave upward (our result) or concave downward in the
absence of dramatic water stress. Given that understanding the ET-VPD
relationship of the one-way plant response is fundamental to
hypothesizing about any feedbacks between the land surface and the
atmosphere, we analyze why our derived ET-VPD relationship is concave
upward, particularly with respect to the exponent of VPD dependence in
uWUE and the Medlyn unified stomatal conductance model, as other
models and empirical results have suggested exponents varying from 1/2
\cite{Leuning_1990, Zhou_2015, Lin_2018}.

By introducing $n$, the exponent of VPD in uWUE, and $m$, the exponent
of VPD in the stomatal conductance model, we can free our theory from
assumptions about VPD dependence:
  \begin{equation}
    g_s = \frac{R \, T}{P} 1.6 \left(1 + \frac{g*}{VPD^m}\right) \frac{*WUE \; ET}{c_a \; VPD^n},
    \label{m_n}
  \end{equation}
where:
\[*WUE = \frac{GPP}{ET}VPD^n,\] and $g*$ is a generic slope parameter
of units $VPD^m$. To determine how the exponent $n$ and $m$ alter the
shape of the ET-VPD dependence we find the roots of the second
derivative of ET, using Eq. (\ref{m_n}) for stomatal conductance
($g_s$), with respect to VPD:

\begin{equation}
\frac{\partial^2 \; ET}{\partial \; VPD^2} = 0 \quad \forall \quad\frac{VPD^m}{g*} = \frac{m \left(m - 2 n - \sqrt{m^{2} - 4 m n + 2 m - 4 n^{2} + 4 n + 1} + 1\right)}{2 n \left(n - 1\right)} - 1.
\label{curves}
\end{equation}

With this result we have defined the family of curves separating
concave up from concave down ET solutions (Fig. \ref{concave}). These
curves are only functions of the exponent of the VPD dependence and a
quantity we call non-dimensional VPD ($VPD^m/g*$). Several important
relations reveal themselves from Eq. (\ref{curves}):

\begin{figure}
  \centering
  \centerline{\includegraphics[width=0.75\textwidth]{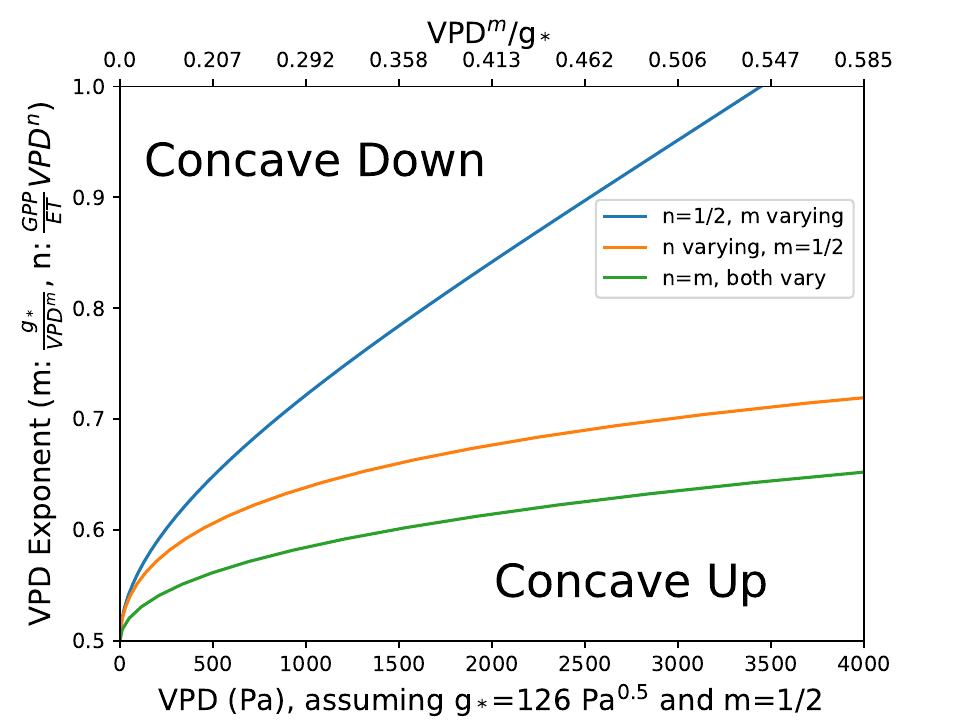}}
  \caption{ Solutions corresponding to inflection points between
    concave up and concave down ET-VPD curves (Eq. (\ref{curves})) for
    three specific scenarios. Solutions are defined in terms of a
    non-dimensional VPD ($VPD^m/g_*$), but to aide physical
    interpretation the horizontal axis is additionally provided in
    terms of dimensionalized VPD assuming $m=1/2$ and
    $g_*=110\; Pa^{1/2}$ (average of all PFT $g_1$). The vertical axis
    has a different interpretation depending on the solution
    curve. For the blue line ($m$ varying), it corresponds to $m$, for
    the orange line ($n$ varying) it corresponds to $n$, and for the
    green line it corresponds to the value of both $n$ and $m$
    ($n=m$). Regions of the parameter space that correspond to concave
    up and concave down results are labeled: for each curve, the
    parameter space ``below'' the curve corresponds to a concave up
    relationship, while the parameter space ``above'' the curve
    corresponds to a concave down relationship.}
  \label{concave}
\end{figure}

\begin{itemize}
  \item For optimal behavior (n, m = 1/2) the ET-VPD curve will be
    concave up regardless of the magnitude of the plant constants
    $g_1$ and uWUE. Therefore, the general concave up nature of our
    results, given an assumption of optimal behavior, is insensitive
    to plant type.
  \item For all physically possible exponents of VPD dependence ($n,
    m$), whether the solution is concave up or concave down does not
    depend on $*WUE$.
  \item In general, increasing the exponent of VPD dependence for
    either $*WUE$ or $g_*$ increases the likelihood of a concave down
    result. Additionally, as the exponent of VPD dependence increases
    from the optimum value of 1/2, whether the curve is concave upward
    or concave downward becomes a function of the plant specific slope
    parameter $g_*$, through non-dimensional VPD
    ($VPD^m/g_*$). Because the exponent of the VPD dependencies is
    capable of altering the fundamental shape of ET-VPD dependence,
    future research investment in understanding the exact VPD
    dependence of stomatal conductance, and further reconciliation of
    empirical and theoretical stomatal and ecosystem behavior should
    be prioritized.
\end{itemize}

While it is possible (and even likely) that in the future some other
form of VPD dependence is derived, at present \cite{MEDLYN_2011} and
\cite{Zhou_2014} established n=m=1/2 as the most likely candidate
given current theory and empirical data. Additionally, we argue that a
concave up result matches physical intuition more than a concave down
result. Plants must maintain nutrient and sugar transport through the
phloem and xylem. To accomplish this, stomata must remain slightly
open \cite{De_2013, Nikinmaa_2013, Ryan_2014}. Furthermore, even if
complete stomatal closure were possible, cuticular water loss and (at
the ecosystem-scale) direct soil evaporation are still sources of ET
which increase with VPD, independent of stomatal closure. Therefore,
in the limit as VPD becomes large and we assume plants are exercising
all strategies to reduce ET, any further increase in VPD should result
in an increase in ET through cuticular water loss and/or direct soil
evaporation. This inevitable transition from conditions when stomata
respond strongly to VPD to conditions when stomata response is
asymptoting towards full closure would cause a concave up ET-VPD
curve, which is matched by our theory. In short, plant response
becomes more limited as VPD increases, while atmospheric demand
monotonically increases with VPD, leading to the result that
atmospheric demand dominates plant response when atmospheric demand is
high.

This analysis allows us to understand the theoretical shape of the ET
response to VPD with environmental conditions held
fixed. Accomplishing this with purely statistical methods applied to
flux observations would be very difficult, given the relatively fast
time scale of plant response and the non-stationarity of (solar
forced) environmental conditions over the relatively coarse (half
hourly) flux estimates (which is required to obtain robust
eddy-covariance statistics). Our results on the shape of the ET-VPD
curve with environmental conditions held fixed can be built upon with
future work examining how changes in VPD and environmental conditions
(e.g. soil water storage) feedback upon one another. In the soil water
storage example, over very long time scales extremely high VPD
perturbations coupled with no precipitation could result in decreases
in soil water storage such that water becomes limiting. This could be
represented by an extension of our framework in which $\lambda$ is
allowed to decrease with decreasing soil water, increasing uWUE and
decreasing g$_1$. Here, we focus our results by framing them with
previous literature to build baseline intuition for ET-VPD dependence.

\section{Conclusions}

We derived a new form of Penman Monteith using the concept of
semi-empirical optimal stomatal regulation \cite{Lin_2015,
MEDLYN_2011} and near constant uWUE \cite{Zhou_2015} to remove the
implicit dependence of stomatal conductance on GPP and ET. With our
new form of Penman Monteith we developed a theory for when an
ecosystem will tend to reduce or increase ET with increasing VPD,
which we framed using previous literature exploring the relationship
between plant parameters, plant types, and climate. The goal was to
understand the range of possible ET responses to VPD and develop some
intuition for how the ET response may vary with plant types and
climate. This intuition can be used to disentangle land atmosphere
feedbacks in more complicated scenarios, and will aid interpretation
of observations and more sophisticated models.

ET response to VPD can vary from strongly water conservative (ET
decreasing in response to increasing VPD), to strongly water intensive
(ET increasing in response to VPD), which is indicative of the
diversity of possible plant water conservation strategies. Higher uWUE
and g$_1$ values increase the likelihood of a positive ET response,
while decreasing temperature and increasing g$_a$ amplify the
magnitude of the response. Previous literature \cite{Zhou_2015,
Lin_2015} suggests that crops, through association with higher g$_1$
and uWUE, are more likely to exhibit a positive ET response to
VPD. Shrubs (lower uWUE), C4 plants (lower g$_1$), gymnosperm trees
(lower g$_1$), and plants in arctic and boreal climates (lower g$_1$)
are more likely to exhibit a negative ET response to VPD. However
interpretation of g$_1$-induced variability is partially muddied by
ambiguity in applying leaf-scale estimations of g$_1$ to the ecosystem
scale \cite{Medlyn_2017}.

Our paper builds intuition for how plants respond to VPD
perturbations. We show that given optimal stomatal function and fixed
environmental conditions, the ET-VPD dependence is theoretically
concave upward, with ET increasing with increasing VPD as VPD
increases past some critical value where $\frac{\partial \;
ET}{\partial \; VPD} = 0$. However future research should focus on
fully understanding the functional form of VPD dependence, as this
concave up result is sensitive to the exponent of VPD dependence,
which we currently believe is 1/2 for both uWUE and the stomatal
conductance model \cite{MEDLYN_2011, Zhou_2014}. Indeed, this
sensitivity to the exponent of VPD dependence is an important result
itself: land surface models, including those used in earth system
models for climate forecasts, employ different assumptions about the
exponent of VPD dependence in stomatal conductance
\cite<e.g.,>[]{Ball_1987, Leuning_1990, MEDLYN_2011}, and these
assumptions can fundamentally change the relationship between ET and
VPD from one that is concave upward (local minimum in ET) to one that
is concave downward (local maximum in ET).

Our results are also applicable to understanding the impact of
expected increases in VPD induced by global change.  Plant
physiological responses to direct CO$_2$ effects
\cite<e.g.,>[]{Swann_2016, Lemordant_2018} receives more attention
than physiological response to indirect effects like increased VPD
\cite{Novick_2016}. Here, we provide a framework for understanding ET
response to VPD using a simplified model of two plant
parameters. Feedbacks between the land and the atmosphere may alter
the net response to a long-timescale global VPD perturbation, but our
focus on the one way plant response to a VPD perturbation in the
atmospheric boundary layer is an important first step to disentangling
such feedbacks, both in observations and model simulations of the
present and future. By removing Penman Monteith's dependence on
implicit relationships between GPP, VPD, and ET, we allow for explicit
future analysis of plant-VPD feedbacks in the atmospheric boundary
layer (Eq. (\ref{et})). Our approach can be extended to examine
varying plant response to more nuanced consideration of plant type and
climate. Any plant physiological heterogeneity or feedback that can be
conceptualized with shifts in $g_1$ \cite<e.g.>[]{Lin_2015,
Medlyn_2017} and/or uWUE \cite<e.g.>[]{Zhou_2014} are representable
within our framework, which opens the door for a hierarchy of more
sophisticated climate- and plant-specific analyses of ET sensitivity
to environmental variables (including VPD). We argue that such
simplified conceptual frameworks are critical tools for disentangling
land-atmosphere feedbacks at various scales, from diurnal to seasonal
and beyond, and to characterize ET response in a warmer,
atmospherically drier, and enriched CO$_2$ world.

\nocite{AT-Neu, AU-ASM, AU-ASM, AU-Cpr, AU-DaP, AU-DaS, AU-Dry, AU-Gin,
  AU-How, AU-Gin, AU-DaP, AU-Tum, AU-Whr, AU-Wom, BE-Lon, BE-Vie,
  BR-Sa3, CA-Qfo, CA-SF1, CA-SF1, CH-Cha, CH-Dav, CH-Fru, DE-Geb,
  DE-Gri, DE-Hai, DE-Gri, DE-Lkb, DE-Seh, DE-Tha, DK-Sor, FI-Hyy,
  FI-Sod, FR-Gri, FR-LBr, IT-Col, IT-Cpz, IT-Lav, IT-MBo, IT-Noe,
  IT-Ren, IT-Ro2, IT-SRo, IT-Tor, NL-Loo, RU-Fyo, US-AR1, US-AR1,
  US-ARM, US-Blo, US-KS2, US-MMS, US-Me2, US-NR1, US-Ne1, US-Ne1,
  US-Ne1, US-SRG, US-SRM, US-Syv, US-Ton, US-Var, US-WCr, US-Wkg,
  ZA-Kru, ZM-Mon}

\acknowledgments This work used eddy covariance data acquired and shared
by the FLUXNET community, including these networks: AmeriFlux, AfriFlux,
AsiaFlux, CarboAfrica, CarboEuropeIP, CarboItaly, CarboMont,
ChinaFlux, Fluxnet-Canada, GreenGrass, ICOS, KoFlux, LBA, NECC,
OzFlux-TERN, TCOS-Siberia, and USCCC. The ERA-Interim reanalysis data
are provided by ECMWF and processed by LSCE. The FLUXNET eddy covariance
data processing and harmonization was carried out by the European Fluxes
Database Cluster, AmeriFlux Management Project, and Fluxdata project of
FLUXNET, with the support of CDIAC and ICOS Ecosystem Thematic Center,
and the OzFlux, ChinaFlux and AsiaFlux offices. We would like to thank
Dr. Trevor Keenan for providing tools for citation of the FLUXNET2015
dataset, available at
\url{https://github.com/trevorkeenan/FLUXNET_citations}. This material is
based upon work supported by the National Science Foundation Graduate
Research Fellowship under Grant No. DGE 16-44869. All code and data used
in this analysis, including those used to generate the figures and
tables, are publicly available at
\url{https://github.com/massma/climate\_et}. We would
like to thank Emma Robinson and one anonymous reviewer for thoughtful
comments that greatly improved the strength of the manuscript.

\bibliography{references}

\end{document}